\documentclass[aps,prd,twocolumn,showpacs,preprintnumbers,amsmath,amssymb]{revtex4}   
\usepackage{graphicx}% Include figure files
\usepackage{dcolumn}% Align table columns on decimal point
\usepackage{bm}% bold math
\usepackage{amsfonts}
\usepackage{amsmath,amssymb}
\newcommand {\beq} {\begin{equation}}
\newcommand {\eeq} {\end{equation}}
\newcommand {\beqa}{\begin{eqnarray}}
\newcommand {\eeqa}{\end{eqnarray}}
\newcommand {\del} {\partial}
\newcommand {\tr}{{\rm tr\,}}

\newcommand {\vev}  [1]{\ensuremath{\langle #1 \rangle}}
\newcommand {\rf}   [1]{(\ref{#1})}
\newcommand {\eexp} [1]{\ensuremath{\mbox{e}^{#1}}}

\begin{document}

%%%%%%%%%%%%%%%%%%%%%%%%%%%%%%%%%%%%%%%%%%%%%%%%%%%%%%%%%%%%%%%%%%%%
%  TITLE / AUTHOR                                                  %
%%%%%%%%%%%%%%%%%%%%%%%%%%%%%%%%%%%%%%%%%%%%%%%%%%%%%%%%%%%%%%%%%%%%

\title{
A general approach to the sign problem\\
---the factorization method with multiple observables
}
 
\author{Konstantinos~N.~Anagnostopoulos$^{1}$}
\email{konstant@mail.ntua.gr}
\author{Takehiro~Azuma$^{2}$}
\email{azuma@mpg.setsunan.ac.jp}
\author{Jun Nishimura$^{3,4}$}
\email{jnishi@post.kek.jp}

%\address{
\affiliation{
$^{1}$Physics Department, National Technical University of Athens,
Zografou Campus, GR-15780 Athens, Greece \\
$^{2}$Institute for Fundamental Sciences, Setsunan University,\\
17-8 Ikeda Nakamachi, Neyagawa, Osaka, 572-8508, Japan \\
$^{3}$KEK Theory Center, High Energy Accelerator Research Organization, 
Tsukuba 305-0801, Japan \\
$^{4}$Department of Particle and Nuclear Physics, 
School of High Energy Accelerator Science,
%\\
Graduate University for Advanced Studies (SOKENDAI),
Tsukuba 305-0801, Japan
%${}^3$Department of Particle and Nuclear Physics, 
%		The Graduate University for Advanced Studies (SOKENDAI),
%			Tsukuba, Ibaraki 305-0801, Japan
}

\date{February, 2011; preprint: KEK-TH-1399
%, hep-th/yymmnnn
%\today %%new
}% It is always \today, today,

%%%%%%%%%%%%%%%%%%%%%%%%%%%%%%%%%%%%%%%%%%%%%%%%%%%%%%%%%%%%%%%%%%%%
%  ABSTRACT    	                                                   %
%%%%%%%%%%%%%%%%%%%%%%%%%%%%%%%%%%%%%%%%%%%%%%%%%%%%%%%%%%%%%%%%%%%%

\begin{abstract}
The sign problem is a notorious problem, which occurs
in Monte Carlo simulations of a system
with the partition function whose integrand is not real positive.
The basic idea of the factorization method
applied on such a system
is to control some observables in order to determine and 
sample efficiently
the region of configuration space
which gives important contribution to the partition function.
We argue that it is crucial to choose appropriately 
the set of the observables to be controlled
in order for the method to work successfully in a general system.
This is demonstrated by an explicit example, in which it 
turns out to be necessary to control more than one observable.
Extrapolation to large system size
is possible due to the nice scaling
properties of the factorized functions,
and known results obtained by an analytic method
are shown to be consistently reproduced.
\end{abstract}

%\pacs{11.25.-w; 11.25.Sq}
%11.25.-w Theory of fundamental strings
%11.25.Sq Nonperturbative techniques; string field theory

%% %%%Specific theories and interaction models; particle systematics
%% 12.38.Gc 	Lattice QCD calculations 
%% (see also 11.15.Ha Lattice gauge theory)
%% 11.15.Ha 	Lattice gauge theory 
%% (see also 12.38.Gc Lattice QCD calculations)

%% %%%statistical physics
%
%% 05.10.-a 	
%% Computational methods in statistical physics and nonlinear dynamics 
%% (see also 02.70.-c in mathematical methods in physics)
%
%% 05.10.Ln 	Monte Carlo methods 
%% (see also 02.70.Tt, Uu in mathematical methods in physics; 
%% for Monte Carlo methods extensively used in subdivisions of physics, 
%% see the appropriate section; for example, see 52.65.Pp in plasma simulation)

%% %%%mathematical methods in physics
%
%%  02.70.-c 	Computational techniques; simulations 
%% (for quantum computation, see 03.67.Lx; 
%% for computational techniques extensively used in subdivisions of physics, 
%% see the appropriate section; for example, see 47.11.-j 
%% Computational methods in fluid dynamics)
%
%% 02.70.Tt 	Justifications or modifications of Monte Carlo methods
% 
%% 02.70.Uu 	Applications of Monte Carlo methods 
%% (see also 02.50.Ng in probability theory, 
%% stochastic processes, and statistics, 
%% and 05.10.Ln in statistical physics)

\maketitle

%%%%%%%%%%%%%%%%%%%%%%%%%%%%%%%%%%%%%%%%%%%%%%%%%%%%%%%%%%%%%%%%%%%%
%  1. INTRODUCTION                                                 %
%%%%%%%%%%%%%%%%%%%%%%%%%%%%%%%%%%%%%%%%%%%%%%%%%%%%%%%%%%%%%%%%%%%%

%\paragraph*{Introduction.---}
\section{Introduction}

Monte Carlo simulation is a powerful tool for studying statistical
systems from first principles.  When the partition function has an
integrand which is not real positive, however, one encounters a
notorious technical problem called the sign problem.  There have been
many proposals, but most of them are successful only for a very
special system \cite{Bietenholz:1995zk} or for a very small region of
the parameter space \cite{Lombardo:2009tf}.  In Ref.~\cite{0108041},
two of the authors proposed a method termed the factorization method,
which is based on the factorization property of the density of states
of properly chosen physical observables. It has been tested in a
random matrix theory for finite density QCD \cite{Ambjorn:2003rr}, and
applied also to the lattice QCD at finite density
\cite{Fodor:2007vv,Ejiri:2007ga} with some important new ideas
\cite{Ejiri:2007ga}.
%In the lattice theory several other methods
%have been proposed and are currently under intensive investigation
%\cite{de Forcrand:2002ci,Fodor:2001au,Aarts:2008rr,Allton:2002zi}.
In the lattice gauge theory, a lot of efforts are actively pursued also
using other methods such as analytic continuation \cite{de
Forcrand:2002ci}, multiparameter reweighting \cite{Fodor:2001au},
complex Langevin dynamics \cite{Aarts:2008rr} and the Taylor expansion
method \cite{Allton:2002zi}.

The basic idea of the factorization method is to control some
observables in order to determine and sample efficiently the region of
configuration space which gives important contribution to the
partition function.  While the previous studies
\cite{0108041,Ambjorn:2003rr,Fodor:2007vv,Ejiri:2007ga} suggest its
potential usefulness in a wider range of applications, the choice of
the observables to be controlled seemed rather arbitrary.  In this
work we argue that it is actually crucial to choose them appropriately
in order for the method to work successfully in a general system.
With this new insight, we consider that the factorization method has
become a very promising approach applicable to any interesting
system that suffers from the sign problem.

\section{Sign problem}
%\paragraph*{Sign problem.---}
Let us consider a general system defined by the partition function
\beq
Z = \int dA \, \eexp{-S_{0}[A] + i \Gamma[A]} \ , \\
\label{partition-fn}
\eeq
where $A$ represents the dynamical variables, and 
$S_0$ and $\Gamma$ represent the real part and the imaginary part
of the action, respectively.
Since the integrand of (\ref{partition-fn}) is not real positive
due to $\Gamma$, one cannot view it 
as a sampling probability in a Monte Carlo simulation.
One way to calculate the expectation value 
of an observable ${\cal O}$ is to use
the reweighting formula
\beq
  \label{eq:ovev}
  \vev{{\cal O}} = \frac
 {\vev{{\cal O}\,\eexp{i\Gamma}}_0}
 {\vev{          \eexp{i\Gamma}}_0}\ ,
\eeq
where the expectation values on the right-hand side are taken
with respect to the phase-quenched model
\beq
 \label{eq:Z0}
Z_0 = \int dA \, \eexp{-S_{0}[A]}\, ,
\eeq
which can be simulated in the usual manner.
The expectation value $\vev{\eexp{i\Gamma}}_0$ is nothing but
the ratio of the partition functions $Z/Z_0$.
Therefore it decreases exponentially for large system size $V$ as 
$\eexp{-V \Delta f}$, where 
$\Delta f>0$ 
represents 
the difference in the free energy density of the two systems.
This can happen due to huge cancellations from $\eexp{i\Gamma}$,
and occurs also in the numerator of (\ref{eq:ovev}).
As a result one needs 
${\rm O}(\eexp{\mbox{\scriptsize const.} V})$ 
configurations
to compute 
the expectation value $\vev{{\cal O}}$
%an observable 
with given accuracy. 
This is called the sign problem.

There is also a general problem called the ``overlap problem''
in
using a reweighting formula like (\ref{eq:ovev}).
This occurs since the region $R_0$ of configuration space one can sample
effectively by simulating the phase-quenched model
has very little overlap with the region $R$ which gives important 
contribution in the evaluation of 
$\vev{{\cal O}\,\eexp{i\Gamma}}_0 $
and $\vev{\eexp{i\Gamma}}_0$.
One has to run a very long simulation until one can sample
enough configurations in the region $R$.

\section{Factorization method}
\label{sec:fac}
%\paragraph*{Factorization method.---}
In order to reduce the overlap problem,
we control some observables so that one can sample configurations
in the region $R$ efficiently.
Let us introduce a set of such observables 
\beq
 \label{eq:oset}
\Sigma = \{  {\cal O}_k \, | \,  k=1, \cdots  , n \}
\eeq
and define the normalized observables
\beq
\widetilde{\cal O}_k = 
\frac{{\cal O}_k}
{\vev{{\cal O}_k}_0} \ .
\label{normalizedO}
\eeq
Their expectation values
can be written as
\beq
\label{eq:ojtilde}
\langle \tilde {\cal O}_j \rangle
= \int  \left[\prod_{k=1}^n dx_k \right] \, x_j \, \rho(x_1 , \cdots , x_n) 
\eeq
using the density of states
\beq
\label{rho_full_gen1}
\rho(x_1 , \cdots  , x_n) = 
%\Big
\left\langle \prod_{k=1}^n
\delta (x_k -\widetilde{\cal O}_{k}) 
%\Big
\right\rangle \ . 
\eeq
Applying the reweighting formula to (\ref{rho_full_gen1}),
one can easily derive the factorization property
\beq
\rho(x_1 , \cdots  , x_n) = \frac{1}{C} \, 
\rho^{(0)}(x_1 , \cdots , x_n) \, w(x_1 , \cdots , x_n) \ ,
\label{fac-prop}
\eeq
where $C=\vev{\eexp{i\Gamma}}_0$ and
\beq
\label{rho_pq_gen1}
\rho^{(0)}(x_1 , \cdots  , x_n) = 
%\Big
\left\langle \prod_{k=1}^n
\delta (x_k -\widetilde{\cal O}_{k}) 
%\Big
\right\rangle_0 
\eeq 
is the density of states for the phase-quenched model.
The correction factor
% $w$
$w(x_1 , \cdots , x_n)$ 
is given by
\beq
w(x_1 , \cdots , x_n)
=\vev{          \eexp{i\Gamma}}_{x_1 ,  \cdots  , x_n }
\label{def-w}
\eeq
as an expectation value 
in 
a constrained system
\begin{equation}
 Z(x_1 ,  \cdots  , x_n) = 
\int dA \, \eexp{-S_{0}} \, 
\prod_{k=1}^n \delta (x_k -\widetilde{\cal O}_{k}) \ . 
\label{ix_pf}
\end{equation}
In what follows we assume that 
$w(x_1 , \cdots , x_n)$
is real and positive \cite{endnote}.

When the system size $V$ goes to $\infty$,
the expectation values are given by 
$\vev{\widetilde{\cal O}_k} = \bar{x}_k$,
where $(\bar{x}_1 , \cdots  , \bar{x}_n)$
denotes the position of the peak of 
$\rho(x_1 , \cdots  , x_n)$.
Hence they can be obtained by solving
% the saddle-point equation
\beq
\label{sp-eq}
f^{(0)}_k (x_1 , \cdots , x_n) 
= - \frac{\del}{\del x_k} \Phi(x_1 , \cdots , x_n) 
\eeq
for $k=1, \cdots , n $, where we have defined 
\begin{alignat}{3}
&f^{(0)}_k (x_1 , \cdots , x_n) 
=
\lim_{V\rightarrow \infty}  \frac{1}{V}
\frac{\del}{\del x_k}
\log
\rho^{(0)}(x_1 , \cdots , x_n)   \ ,
\nonumber
\\
&\Phi(x_1 , \cdots , x_n) 
= \lim_{V\rightarrow \infty}  \frac{1}{V}
\log w(x_1 , \cdots , x_n)  \ .
\label{Phi-def}
\end{alignat}
Note that the right-hand side of (\ref{sp-eq}),
which represents the effect of $\Gamma$,
can largely shift the peak from that of
$\rho^{(0)}(x_1 , \cdots , x_n)$, which is given by
$x_k = 1$ due to the chosen normalization (\ref{normalizedO}).
Thus we can obtain
% the expectation values
$\vev{\widetilde{\cal O}_k}$
including the contribution from 
%important 
configurations that are difficult
to sample by simulating the phase-quenched model 
without any constraints.
%% Thus the simulation of the constrained system (\ref{ix_pf})
%% enables us to sample important configurations that are difficult
%% to sample by simulating the phase-quenched model 
%% without any constraints.
%The left-hand side of (\ref{sp-eq}) can be also calculated
%by simulating 

\section{Choice of observables}
\label{sec:multiobs}
%\paragraph*{Choice of observables.---}

What is written above is mathematically correct for
an arbitrary choice of the set $\Sigma$
of observables.
%A natural question that arises is 
We will argue, however, that the overlap problem can still occur 
in the evaluation of
(\ref{def-w})
by simulating (\ref{ix_pf}).
Note that the only difference from
evaluating the denominator of (\ref{eq:ovev})
is the existence of the constraints.
In fact it turns out to be important to choose the set $\Sigma$
appropriately so that the overlap problem is removed.
%
%The reasons for the appearance of the 
%In order to reduce the overlap problem, 
%one should add a new observable in the set $\Sigma$.

%In order to see how this idea works,
In order to clarify the 
overlap problem in the evaluation of (\ref{def-w}),
%and to find a possible cure to it,
let us consider another observable ${\cal O}_{n+1}$,
and define the corresponding
functions 
$\rho^{(0)}$ and $w$
with $n$ replaced by $n+1$.
Then we obtain the relation
\begin{alignat}{3}
& w(x_1 , \cdots , x_n)  \nonumber\\
& = \frac{
\int d x_{n+1} \, \rho^{(0)}(x_1 , \cdots , x_{n+1}) 
\,  w(x_1 , \cdots ,  x_{n+1})  }{
\int d x_{n+1} \,  \rho^{(0)}(x_1 , \cdots , x_{n+1}) 
} \ .
\label{w-relations}
\end{alignat}
When one simulates (\ref{ix_pf}), one mostly samples configurations
with $\widetilde{\cal O}_{n+1}$
close to $\langle \widetilde{\cal O}_{n+1} \rangle_{x_1 , \cdots , x_n}$.
However, it can happen that 
%$w(x_1 , \cdots , x_{n+1})$ 
the integration over $x_{n+1}$
in the numerator of (\ref{w-relations}) has important contribution
from the region of $x_{n+1}$ not close to
$\langle \widetilde{\cal O}_{n+1} \rangle_{x_1 , \cdots , x_n}$
%due to the $x_{n+1}$-dependence of $w(x_1 , \cdots , x_{n+1})$.
% is large. 
if $w(x_1 , \cdots , x_{n+1})$
has strong dependence on $x_{n+1}$.
In that case, one clearly has some remaining overlap problem 
in obtaining $w(x_1 , \cdots , x_n)$ by simulating (\ref{ix_pf}).
%This clearly implies that the overlap problem still remains.
The overlap problem can be reduced further
%In that case 
by including the observable ${\cal O}_{n+1}$
in the set $\Sigma$.
One can, in principle, repeat this procedure until 
the overlap problem is totally removed.

%Let us next consider in the opposite way.
%Namely we 
Let us then discuss what is the minimal set of observables
that is sufficient to remove the overlap problem.
%For practical purposes we need to find such $\Sigma$
%with a reasonably small number of observables.
For that, let $\Sigma$ be a set 
%which includes 
of all possible observables, 
and assume that there is no more overlap problem.
%In order to address this issue, we assume that
%there is no more overlap problem for a set $\Sigma$,
%
%Typically the set $\Sigma$ includes many observables,
%and we need to 
%and discuss which operators can actually be omitted.
Here we consider a simplified situation, in which 
the right-hand side of (\ref{sp-eq}) is nonzero for 
$1 \le k \le K$, and vanishes for the rest; \emph{i.e.,}
\beq
\frac{\del}{\del x_j}\Phi(x_1 , \cdots , x_n)  = 0 \quad
\mbox{for~$K<j\le n$} \ 
\label{Phi-prime-zero}
\eeq
near the peak $(\bar{x}_1 , \cdots , \bar{x}_n)$.
Then we can actually use a smaller set
$\Sigma ' = \{  {\cal O}_k \, | \,  k=1, \cdots  , K \} $
without having the overlap problem in evaluating 
$w(x_1 , \cdots , x_K)$ around the peak.
Furthermore, from (\ref{Phi-prime-zero})
one can easily show that the expectation value
$\vev{\widetilde{\cal O}_j}$ for $j > K$ can be evaluated as
$ \langle \widetilde{\cal O}_j \rangle
\simeq 
\vev { \widetilde{\cal O}_j }_{\bar{x}_1 , \cdots , \bar{x}_K}$.
That this happens in spite of the sign problem 
can be understood by noting that 
\beq
\langle \widetilde{\cal O}_j \, \eexp{i \Gamma}
 \rangle_{\bar{x}_1 , \cdots  , \bar{x}_K}
\simeq 
\vev { \widetilde{\cal O}_j }_{\bar{x}_1 , \cdots , \bar{x}_K} 
\langle \eexp{i \Gamma}
\rangle_{\bar{x}_1 , \cdots , \bar{x}_K} \ ;
\eeq
namely the observables $\widetilde{\cal O}_j$ ($j >K$)
are decorrelated with $\eexp{i \Gamma}$.
The fluctuation of the phase plays an important role
in the determination 
of $(\bar{x}_1 , \cdots  , \bar{x}_K)$ through
the saddle-point equation (\ref{sp-eq}),
but once they are determined, 
the phase can be neglected completely
when evaluating the expectation values of
the observables $\widetilde{\cal O}_j$ ($j >K$).

In general, we can expect
(\ref{Phi-prime-zero}) to hold only approximately.
In that case the systematic error involved in
the above evaluation of 
$\vev{\widetilde{\cal O}_j}$
is given by ($j,l > K$) 
\beqa
 \Delta \bar{x}_j &=& \left.
({\cal H}^{-1})_{jl} 
\frac{\del }{\del x_l} \Phi(x_1 , \cdots , x_n) 
\right|_{\bar{x}_1 , \cdots  , \bar{x}_n} \ ,
\label{sys-error} \\
{\cal H}_{jl} &\equiv & \left. 
\frac{\del}{\del x_l} 
f_j^{(0)}(x_1 , \cdots , x_n) 
%\frac{\del^2}{\del x_j \del x_k} 
%\log \rho^{(0)}(x_1 , \cdots , x_n) 
\right|_{\bar{x}_1 , \cdots  , \bar{x}_n}  \ .
\eeqa
This may also cause a small overlap problem
in evaluating $w(x_1 , \cdots , x_K)$ around the peak.

%From these discussions, it is clear that we need to
%choose the set $\Sigma$ so that the overlap problem
%is maximally reduced.
%is reduced to a practically acceptable level.
%Let us discuss how to choose the set $\Sigma$ of observables.
%What one can do is to 
A practical way to search for the observables to be included
in the set $\Sigma$ is to
calculate an ensemble average of $\eexp{i \Gamma}$ 
in the phase-quenched model with an observable $\widetilde{\cal O}$
fixed to $x$.
If it becomes much larger
%close to 1
in some region of $x$, 
the observable should be considered as a candidate.
We expect that there are many systems in which
only a few observables have to be included in the set $\Sigma$.

%%%%%%%%%%%%%%%%%%%%%%%%%%%%%%%%%%%%%%%%%%%%%%%%%%%%%%%%%%%%%%%%%%%%
%  2. SIMULATION TECHNIQUES                                        %
%%%%%%%%%%%%%%%%%%%%%%%%%%%%%%%%%%%%%%%%%%%%%%%%%%%%%%%%%%%%%%%%%%%%

\section{An explicit example}
%\paragraph*{An explicit example.---}

Let us demonstrate how the method works in an explicit example.
Here we study a matrix model defined by the partition
function \cite{0108070}
\beqa
\label{z_int_out}
Z &=& \int dA \, \eexp{-S_{\rm b}} \, (\det {\cal D})^{N_{\rm f}} \ ,  \\
S_{\rm b} &=& \frac{1}{2} \, N \, 
\sum_{\mu=1}^4 \tr (A_{\mu})^{2} \ , \quad
{\cal D} = \sum_{\mu=1}^4 \Gamma_{\mu} \otimes A_{\mu}  \ , 
\label{Sb-D-def}
\eeqa
where $A_{\mu}$ $(\mu= 1,\cdots,4)$
are $N \times N$ Hermitian matrices
and the $2\times 2$ matrices $\Gamma_{\mu}$
are Pauli matrices $\Gamma_j= \sigma_j$ for $j=1,2,3$,
and $\Gamma_4 = i {\bf 1}$.
The system has a rotational SO(4) symmetry corresponding to 
$A_\mu \mapsto O_{\mu\nu} A_\nu$ with $O \in \mbox{SO}(4)$.
The determinant $\det {\cal D}$ is complex in general. 
Under parity transformation $A_4\to -A_4$, $A_j\to A_j$ ($j
\ne 4)$, it transforms as 
$\det{\cal D}\to (\det{\cal D})^*$. 
This implies that $\det{\cal D}$ is real
for configurations with $A_4=0$. 
{}From this fact alone, it follows that 
the phase of the determinant becomes stationary
for configurations with $A_4=A_{3}=0$
since one cannot have a phase fluctuation 
within a linear perturbation around 
such configurations \cite{Nishimura:2000ds}.

This model was proposed \cite{0108070} as a toy model for the
spontaneous symmetry breaking (SSB) of the SO(10) rotational symmetry
expected to occur in the IKKT model, a conjectured
nonperturbative formulation of superstring theory \cite{Ishibashi:1996xs}.
In the IKKT model, the space-time is represented by the
eigenvalue distribution of $A_\mu$ $(\mu= 1,\cdots,10)$
\cite{9802085,endnote-spt}, and
the SSB of SO(10)
%breaking of the SO(10) rotational symmetry 
realizes a scenario for
{\it dynamical} compactification of extra dimensions in superstring
theory. This scenario is supported by explicit calculations based on
the Gaussian expansion 
method (GEM) \cite{Nishimura:2001sx,higher,Aoyama:2010ry}.
It is considered \cite{Nishimura:2000ds}
that the SSB is induced by the phase of the
complex Pfaffian obtained 
by integrating out the fermionic variables.
%is complex, and its phase is expected to 
%induce the SSB \cite{Nishimura:2000ds}.

In the model  (\ref{z_int_out}), the $\textrm{SO}(4)$ rotational symmetry
is expected to be
spontaneously broken in the large-$N$ limit with $r=N_{\rm f}/N$
fixed, where $N_{\rm f}$ is the exponent in (\ref{z_int_out}).  As an
order parameter, we consider the ``moment of inertia tensor'' 
\beq
\label{tmunudef}
T_{\mu  \nu} = \frac{1}{N} \tr (A_{\mu} A_{\nu})\, ,
\eeq 
and its real positive
eigenvalues $\lambda_{k}$ ($k=1,\cdots, 4$) ordered as $ \lambda_{1}
\geq \lambda_{2} \geq \lambda_{3} \geq \lambda_{4} $. 
If their expectation values turn out to be unequal in the large-$N$ limit,
it implies the SSB of the SO(4) symmetry.

The model (\ref{z_int_out}) was studied by the GEM up to the ninth
order for $r\le 2$ \cite{0412194}. It was found that the
$\textrm{SO}(4)$ symmetry is spontaneously broken down to
$\textrm{SO}(2)$. 
Furthermore, by controlling the eigenvalues to have a hierarchy
$\lambda_d \gg \lambda_{d+1}$, one can obtain
$d$-dimensional configurations ($d = 1,2,3$), for which 
the phase fluctuations become milder according 
the arguments below Eq.~\protect\rf{Sb-D-def}.
%This, together with the property that the size of
%phase fluctuations can be controlled \cite{noteondim}
%by the magnitude of the eigenvalues
%$\lambda_{1},\ldots,\lambda_{4}$, 
These properties of the model
makes it an ideal testing ground 
for the ideas in Secs.~\ref{sec:fac} and \ref{sec:multiobs}. 
%% The complex action problem turns out to be quite
%% severe despite the simpleness of the model. It is encouraging that,
%% even in this case, an approach like the one proposed in this work can
%% produce meaningful results and provide insightful information about
%% the non-trivial effects of the phase \cite{noteonscales}.

At $r=1$, for instance, the results obtained by GEM are
$\vev{\lambda_1} = \vev{\lambda_2} \simeq  2.1$, 
$\vev{\lambda_3} \simeq  1.0$,
% and 
$\vev{\lambda_4} \simeq 0.8$,
whereas $\vev{\lambda_k}_0 = \frac{3}{2}$ for the phase-quenched model.
Therefore, the expectation values of observables normalized as
(\ref{normalizedO}) are
\beq
\vev{\tilde{\lambda}_1} = \vev{\tilde{\lambda}_2} \simeq  1.4 \ , \quad
\vev{\tilde{\lambda}_3} \simeq  0.7 \ , \quad
\vev{\tilde{\lambda}_4} \simeq 0.5 \ .
\label{GEMprediction}
\eeq

\section{Monte Carlo simulation}
%\paragraph*{Monte Carlo simulation.---}

We study the model (\ref{z_int_out}) at $r=1$, where 
the sign problem is severe.
Since we know that the eigenvalues $\lambda_k$
have strong correlation with the fluctuation of $\Gamma$,
we use ${\cal O}_k = \lambda_k$ ($k=1,\cdots,4$)
as the observables in the set $\Sigma$.
Searching for the solution to \rf{sp-eq} in its full generality is a
formidable task. 
Here we simply
check that the GEM result (\ref{GEMprediction})
is indeed a solution to 
the saddle-point equation (\ref{sp-eq}).
% (\ref{master_eq_so2_0}).
This is sufficient for demonstrating
%The fact that our results are consistent with the GEM suffices to show
that the method is error free and that the generalization of the
factorization method proposed in this work is necessary in order to
remove the remaining overlap problem. 
The question whether one can successfully determine 
the absolute maximum of $\rho(x_1,\ldots,x_4)$ is 
subject to a more detailed investigation.
%() \cite{noteonsol}.

%Therefore
%Instead we will look for solutions consistent with the GEM
%results and 
Let us therefore assume that the SO(2) symmetry remains,
which implies $x_1 = x_2$.
%and search for a solution to the saddle-point equation (\ref{sp-eq})
%with $x_1 = x_2$.
Equation  (\ref{sp-eq}) then reduces to
\begin{alignat}{3}
&  \frac{\partial}{\partial x_k} 
\log \rho^{(0)}_{\textrm{SO(2)}} (x_2,x_3,x_4) = 
- \frac{\partial}{\partial x_k} \log w_{\textrm{SO(2)}} (x_2,x_3,x_4) 
\label{master_eq_so2_0}
\end{alignat}
for $k=2,3,4$, where we have defined
\begin{alignat}{3}
&  \rho^{(0)}_{\textrm{SO(2)}} (x,y,z) = \rho^{(0)} (x,x,y,z) \ , 
\nonumber \\
&  w_{\textrm{SO(2)}} (x,y,z) = 
w(x,x,y,z) \ .
\end{alignat}
In order to be brief \cite{future}, let us only discuss 
Eq.~(\ref{master_eq_so2_0}) for $k=2$. We set $x_{3}=0.7$ and
$x_{4}=0.5$, and solve 
the equation
%eq.~(\ref{master_eq_so2_0}) 
for $x_2$. 
We also define $ w(x) = w_{\textrm{SO(2)}} (x,0.7,0.5)$ and 
$\rho^{(0)}(x) = \rho^{(0)}_{\textrm{SO(2)}} (x,0.7,0.5)$.

\begin{figure}[htb]
\begin{center}
\includegraphics[height=6cm]{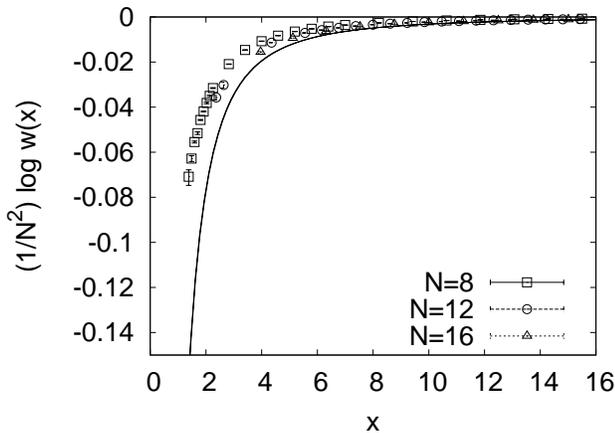}
\end{center}
\caption{The function
$\frac{1}{N^2} \log w(x)$ is plotted for $N=8,12,16$.
%against $x$. 
The solid line represents the asymptotic behavior (\ref{w_2d_x2})
with the coefficients extrapolated to $N=\infty$.  
}
\label{result_so2_2}
\end{figure}
%%%%%%%%%%%%%%%%%%%

\begin{figure}[htb]
\begin{center}
\includegraphics[height=6cm]{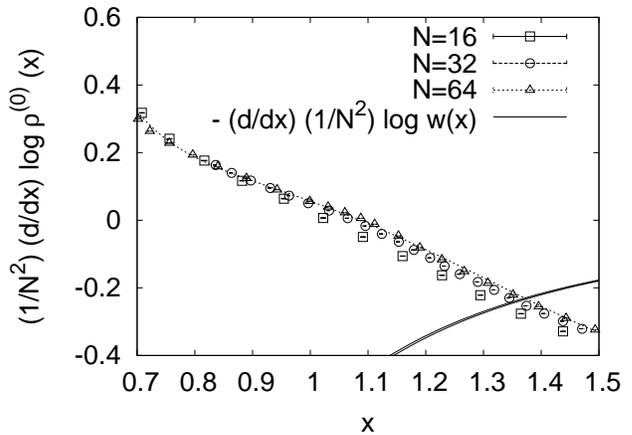}
%% \rotatebox{-90}{
%% \includegraphics[height=6cm]{f-naught-i34_3_r1_a.eps}}
\end{center}
\caption{The function 
$\frac{1}{N^2} \frac{d}{dx}
\log \rho^{(0)}(x)$ is plotted for $N=16,32,64$.
The dashed line is drawn to guide the eye.
We also plot 
$-  \frac{d}{d x} \lim_{N\rightarrow \infty} 
\{ \frac{1}{N^2}\log w(x) \} $
obtained from Fig.~\ref{result_so2_2},
where the two solid lines show the margin of error. 
The value of $x$ at the intersecting point 
is consistent with the GEM result
$\langle \tilde{\lambda}_2 \rangle \simeq 1.4$.
}
\label{result_so2_2f0}
\end{figure}
%%%%%%%%%%%%%%%%%%%

In Fig.~\ref{result_so2_2} we plot our Monte Carlo data for
$\frac{1}{N^2} \log w(x)$. 
It approaches zero at large $x$,
where the dominant configurations have 
$\lambda_2 \gg \lambda_3$,
and the phase $\Gamma$ becomes stationary due to the argument
below Eq.~(\ref{Sb-D-def}). One can actually show that
the asymptotic behavior of $w(x)$ at large $x$ is
\begin{eqnarray}
 \frac{1}{N^2} \log w(x)
\simeq - c_{1} x^{-2} + c_{2} x^{-5/2} \ .
\label{w_2d_x2}
\end{eqnarray}
Fitting our data
to (\ref{w_2d_x2}) for $N=8,12,16$,
we find that finite $N$ effects in the coefficients 
$c_{1}$ and $c_{2}$
are consistent with O($1/N$).
Making extrapolations to $N=\infty$ based on this observation,
we obtain
$c_{1} = 0.322(2)$
and
$c_{2} = 0.021(1)$.
The solid line in Fig.~\ref{result_so2_2} 
represents
Eq.~(\ref{w_2d_x2}) with these extrapolated values.
% coefficients extrapolated to $N=\infty$.
Figure \ref{result_so2_2f0} shows that
the solution is $x = 1.373(2)$,
which is consistent with the GEM result 
$\langle {\tilde \lambda}_{2} \rangle \simeq 1.4$.

\begin{figure}[htb]
%\begin{figure}[t]
\begin{center}
\includegraphics[height=6cm]{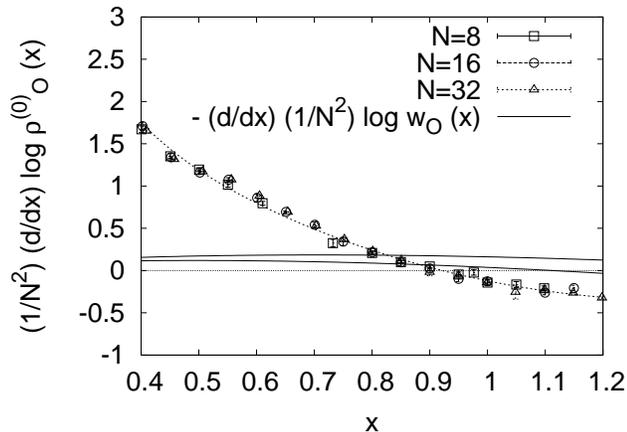}
%% \rotatebox{-90}{
%% \includegraphics[height=6cm]{f-naught-i34_3_r1_a.eps}}
\end{center}
\caption{
The function 
$\frac{1}{N^2} \frac{d}{dx} \log \rho^{(0)}_{{\cal O}}(x)$ 
is plotted for $N=8,16,32$. 
The dashed line is drawn to guide the eye.
% to the maximum of $\rho^{(0)}_{\cal O}(x)$.
% at the vanishing point of $f^{(0)}_{\cal O}(x)$.
We also plot $- \frac{d}{dx} \lim_{N\rightarrow \infty} 
\{ \frac{1}{N^2} \log w_{{\cal O}}(x) \} $, 
where the two solid lines show the margin of error. 
From this figure, we find that the position of the peak 
shifts from $x=0.92$ to $x=0.85(3)$ 
due to the effect of the phase.
}
\label{result_so2_trF2}
\end{figure}

Let us then consider what happens if we add
\beq
\label{trf2} 
{\cal O}= - \frac{1}{N} \sum_{\mu \neq \nu}\tr [A_\mu , A_\nu]^2
%\, ,
\eeq
as the fifth observable in the set $\Sigma$.
We define the corresponding functions $\rho^{(0)}$ and $w$
with five arguments, and also define the reduced functions
\beqa
\rho^{(0)}_{\cal O}(x) &=& \rho^{(0)}(1.4,1.4,0.7,0.5,x)  \ , 
\nonumber \\
w_{\cal O}(x) &=& w(1.4,1.4,0.7,0.5,x)  \ ,
\eeqa
which correspond to fixing 
$\tilde{\lambda}_k$ to the GEM result (\ref{GEMprediction}),
and $\widetilde{\cal O}={\cal O}/\vev{{\cal O}}_0$ to some value $x$.
We find that
$\frac{1}{N^2} \log w_{\cal O}(x)$ approaches zero 
as $x\rightarrow 0$ \cite{endnote2} with the asymptotic behavior
\beq
\label{otildescal}
\frac{1}{N^2} \log w_{\cal O}(x)=
  -d_1 x^{2} + d_2 x^{5/2} \  .
\eeq
Therefore, the observable ${\cal O}$ may, in principle, be a ``dangerous''
one which must be included in the set $\Sigma$.
%(\ref{setSigma}).
Figure \ref{result_so2_trF2} is a plot obtained
similarly to Fig.~\ref{result_so2_2f0},
which shows that the effect of the phase 
represented by (\ref{sys-error})
is to shift the estimate of 
$\langle \widetilde{\cal O} \rangle$
%$\langle {\cal O}\rangle$ 
by $\Delta \bar{x} = 0.07(3)$.
On the other hand, the standard deviation of 
the distribution $\rho^{(0)}_{\cal O}(x)$ is estimated
as $\sigma \sim 0.7/N$ from the slope of the function
plotted in Fig.~\ref{result_so2_trF2} around $x\sim 0.92$.
This means that the deviation $\Delta \bar{x}$ is 
$\lesssim 2 \, \sigma$ for $N \le 16$.
%$\lesssim 10\mbox{\%}$.
%% This is contrary to the observed large effect of
%% the phase on expectation values of $\tilde{\lambda}_k$.
%% %which cannot be omitted from the set $\Sigma$.
%% This can, in principle, cause
%% some overlap problem in obtaining the data 
%% in fig.~\ref{result_so2_2}.
Thus, the remaining overlap problem in obtaining the data 
in Fig.~\ref{result_so2_2} is practically small
as far as this observable (\ref{trf2}) is concerned.
The fact that we are able to reproduce
the GEM result with only four observables $\lambda_k$
in the set $\Sigma$
suggests that the remaining overlap problem is 
indeed not so severe.
%% without including the observable ${\cal O}$ in the set $\Sigma$,
%% however, suggests that the remaining overlap problem is 
%% actually not so severe.
%
%% Thus we consider that there is only 
%% a small overlap problem left in obtaining the data 
%% in fig.~\ref{result_so2_2}.
%% %which was obtained without including the observable ${\cal O}$
%% %in the set $\Sigma$.
%% This is consistent with the fact that we are able to reproduce
%% the GEM result without 
%% including the observable ${\cal O}$ in the set $\Sigma$.

%%%%%%%%%%%%%%%%%%%

%%%%%%%%%%%%%%%%%%%

%%%%%%%%%%%%%%%%%%%%%%%%%%%%%%%%%%%%%%%%%%%%%%%%%%%%

%%%%%%%%%%%%%%%%%%%%%%%%%%%%%%%%%%%%%%%%%%%%%%%%%%%%%%%%%%%%%%%%%%%%
%  Conclusion and Discussions                                      %
%%%%%%%%%%%%%%%%%%%%%%%%%%%%%%%%%%%%%%%%%%%%%%%%%%%%%%%%%%%%%%%%%%%%
\section{Summary and discussion}
%\paragraph*{Summary and discussion.---}
 
In this work we have discussed a general approach
to systems with the sign problem based on the factorization 
property (\ref{fac-prop}) of the density of states.
The method aims at reducing the overlap problem by controlling
the observables which have strong correlation with the 
phase $\eexp{i\Gamma}$. 
Once the solution to the saddle-point equation (\ref{sp-eq}) is
obtained for such observables, all the other observables can be 
investigated without the sign problem.
We have presented an explicit example,
in which Monte Carlo data suggests that 
one has to control only a few observables to remove
the overlap problem almost completely.
We speculate that this is the case in many interesting systems.
Finding the minimal set of observables for each system
as described in Section \ref{sec:multiobs} 
%would constitute an interesting future work.
would be the subject of a future investigation.
%consists a new project, which is beyond the scope of this work.

The main task of the method is to calculate the function (\ref{def-w})
by simulating (\ref{ix_pf}),
which still suffers from cancellations due to $\eexp{i\Gamma}$.
However,
near the solution of the saddle-point equation (\ref{sp-eq}), 
the fluctuation of $\Gamma$ becomes milder than 
in the phase-quenched system without constraints.
Hence it is expected in many cases that
one can compute the function (\ref{def-w}), 
directly or by using asymptotic behaviors 
such as (\ref{w_2d_x2}), for reasonable system size. 
Then the scaling properties 
represented by (\ref{Phi-def})
enable extrapolations to infinite system size.
We hope that Monte Carlo studies of many interesting systems 
that are hindered by the sign problem can be made possible 
by using the factorization method.

%%%%%%%%%%%%%%%%%%%%%%%%%%%%%%%%%%%%%%%%%%%%%%%%%%%%%%%%%%%%%%%%%%%%
%  ACKNOWLEDGEMENTS                                                %
%%%%%%%%%%%%%%%%%%%%%%%%%%%%%%%%%%%%%%%%%%%%%%%%%%%%%%%%%%%%%%%%%%%%

%\hspace{1cm}

%\begin{acknowledgments}

%\section*{Acknowledgments}
  \begin{center}
  {\bf\small ACKNOWLEDGMENTS}
  \end{center}

We would like to thank T.~Aoyama, M.~Hanada, N.~Kawashima and M.~Oshikawa
for useful discussions and comments. 
Computations have been carried out on PC clusters
at KEK and NTUA.
The work of K.N.A.\ was partially funded 
by a PEVE2010 grant from NTUA.
The work of J.N.\ is supported in part by Grant-in-Aid 
for Scientific Research 
(No.\ 19340066 and 20540286)
%and No.\ 20340048(B) for T.Y.\
from Japan Society for the Promotion of Science.
%from the Ministry of Education, 
%Science, and Culture of Japan. 
%\end{acknowledgments}

%\section{Acknowledgments}
%\paragraph*{Acknowledgments.---}

%%%%%%%%%%%%%%%%%%%%%%%%%%%%%%%%%%%%%%%%%%%%%%%%%%%%%%%%%%%%%%%%%%%%
%  REFERENCE                                                      %
%%%%%%%%%%%%%%%%%%%%%%%%%%%%%%%%%%%%%%%%%%%%%%%%%%%%%%%%%%%%%%%%%%%%

%\end{references}


\begin{thebibliography}{99}
\bibitem{endnote2}
As $x$ decreases, the dominant configurations have the property
$[A_\mu , A_\nu ] \approx 0$, meaning that $A_\mu$ are simultaneously 
diagonalizable, \emph{e.g.} as $A_\mu = \mbox{diag}(\alpha_\mu^{(1)},
\cdots , \alpha_\mu^{(N)})$.
For such configurations, the determinant becomes
$\det {\cal D}= \prod_{i=1}^{N} 
\{ \sum_\mu (\alpha_\mu^{(i)}) ^2 \} \ge  0$.
\end{thebibliography}
\end{document}